# Optical control of azo-polymer film loaded surface plasmon-polariton wave

D.G.Zhang[1,3], X.-C.Yuan[2a], G.H.Yuan[1,3], K.J. Moh[1], J.Bu[2], P.Wang[3], L.P.Du[1], J.Lin[1], H.Ming[3]

[1]*School of Electrical and Electronic Engineering, Nanyang Technological University, Nanyang Avenue, Singapore 639798*

[2]*Institute of Modern Optics, Key Laboratory of Optoelectronic Information Science & Technology, Ministry of Education of China, Nankai University, Tianjin 300071, China*

[3]*Department of Physics, University of Science and Technology of China, Hefei, Anhui, 230026, P. R. China*

An active control method of azo-polymer loaded surface plasmon-polaritons (SPPs) wave with tightly focused 532nm laser is proposed and demonstrated in this paper. Theoretical analysis and experimental results are consistent, which confirm the validity of this method. The optically active control method has the advantage of simple samples preparation, high spatial resolution and selectively control of the SPPs wave. This optical control of SPPs wave method has potential applications in integrated plasmonics.

[a] Author to whom correspondence should be addressed. Email: xcyuan@nankai.edu.cn

Surface plasmon-polaritons (SPPs), originating from the resonance interaction between the surface charge oscillation and the electromagnetic filed of the light open the perceptively new filed in optoelectronics[1]. In the last years there has been an increased effort toward the development of SPPs-based devices, and the design of appropriate SPPs waveguides which constitutes a key element in this route [2-4]. One of the currently hotly investigated configurations relies on the extension of the concept of dielectric-loaded waveguides to plasmon polaritons mode which is fabricated by the deposition of a dielectric stripe on top of a metallic surface [4-7]. This approach has been experimentally demonstrated and the theoretical analysis of the properties of these dielectric-loaded SPP waveguides (DLSPPWs) shows their high potential in integrated photonics. In fact, bends, splitters, and couplers based on these DLSPPW have already been experimentally proved.[8, 9] Moreover, the dielectric layer can be artificially doped by suitable materials so that its optical properties could be externally modified to actively control the SPP mode. This configuration opens then an interesting way for active plasmonics, e.g., electrically or optically controlled plasmonics[2], such as the Mach-Zehnder interferometer proposed in Ref [8]. To realize active controlling, proper active dielectric materials should be used to load the surface plasmons. In our experiments, azobenzene (AZO) containing polymer is selected.

The azobenzene (AZO) containing polymer is one of the most promising materials, which play both roles as mesogens and photosensitive chromophores, and has been widely explores in holographic gratings, optical data storage, far-field optical lithography beyond diffraction limit, and bottom-up nanotechnology[10]. Most of the applications are based on photoisomerization between a *trans*-AZO and a *cis*-AZO state. Accompanying with the photoisomerization, photochemical phase transition might be acquired since a trans-AZ could be a liquid crystal (LC) mesogen, because the molecular shape is rod-like, whereas a *cis*-AZO never shows any LC phase owing to its bent shape. Upon irradiation, *tran*-AZO transfers to *cis*-AZO and then transfers back to trans-AZO state because that *cis*-AZO state is not a stable state. As a results, photoinduced alignment of AZO mesogens with their transition moments almost perpendicular to the polarization direction of the incident light has been obtained according to the Weigert effect, leading to a large photoinduced anisotropy[11, 12]. Another property of this molecule is that the photo-induced anisotropy can be erased under irradiation of circular polarized laser [13]. These two properties open one door for active control of SPPs wave with the azobenzene molecules, which is optical controlling the refractive index of the azo-polymer and then induce the control of the wave-vector of the azo-polymer loaded SPPs wave. In this paper, we theoretically analyze and experimentally demonstrate the optical control of the azobenzene containing polymer films loaded SPPs wave which is excited by the radial polarized laser beam.

The azobenzene containing polymer used in this work is Poly [(methyl methacrylate)-co-(Disperse Red 1 acrylate)] from SIGMA-ALDRICH which is dissolved in cyclopentanone solution. The mixed solution is spinning coated onto 50nm gold film deposited on glass substrate with E-beam evaporator. Then the obtained azo-polymer film is baked for 10min at 105$^{o}$C to remove the solvent. The reflective indexes of azo-polymer film are 1.5537, 1.605 at 780nm and 532nm respectively, and the thickness is about 54nm, which are measured with the J.A.Woollam Inc. HS-190 variable angle ellipsometer.

The experimental set-up is shown in Fig.1. Two laser beams (532nm and 780nm) are aligned to propagating through one line. The SPPs waves are excited with radial polarized 780nm laser, which exhibits axially symmetric polarizations over their cross sections and thus the entire beam is p-polarized with respect to the interface. The detail of the radial polarizer laser generating method can be find in Ref[14]. The linear or circular polarized 532nm laser is used to active control of the SPPs waves by generating and erasing the birefringence of the azobenzene molecules, which is in the absorption band of the azobenzene polymer films. The power of the 532nm laser before enter into the objective is about 2mw. The two laser beams are expanded with lens assembles to overfill the back aperture of the oil-immersed objective (60X) which has an N.A. of 1.42 corresponding to an angle range from -69.5° to 69.5° [15]. We use a strong overfilling of the back aperture, corresponding to a filling factor of f≈2. The same objective is used both for generating the azo-polymer film loaded SPPs wave and focusing 532nm onto the film at the same place. The Fourier plane of the leaky radiation image of SPPs wave is collected with one CCD camera (Thorlab DCU 224M, 1280x1024 resolution). A 532 nm long pass edge filter is placed before the CCD to filter the incident 532 nm laser out.

Because the photo-induced anisotropy of azobenzene molecules is strongly related with the polarization of excited laser, it is necessary to investigate the focusing properties of the linear or circular polarized 532nm laser by a high N.A objective lens, which can be numerical simulated based on the diffraction theory [16-18]. Fig.2 is the calculated two-dimensional distribution of $|E|^2$ on the focal plane (it is also the plane of the azo-polymer film) for a highly focused linear polarized 532nm laser. Fig.2 (a, b, c) are the transversal component $|E_\theta|^2$, $|E_r|^2$ and longitudinal component $|E_z|^2$ of the focused linear polarized laser (the polarization is along horizontal axis). It is clear shown in Fig.2 (a, b, c) (the right color bar) that $|E_\theta|^2$ is very smaller than $|E_r|^2$ and $|E_z|^2$, so $|E_\theta|^2$ can be neglected in the reorientation process of the trans-azobenzene molecules. The $E_z$ component is perpendicular to the focal plane, so the transition moment (long axis) of the rods like trans-azobenzene will re-orientate in the focal plane and finally fall perpendicular to the radial direction ($E_r$) under irradiation of the focused laser beam. Fig.2 (d) displays the reorientation of the trans-azobenzene with their long axis perpendicular to the electric vector ($E_r$) in the focal plane. The lines represent the electric vector ($E_r$) in the focal plane and the length represent the intensity of the field. The field intensity of $|E_r|^2$ and $|E_z|^2$ are strong in the polarization direction (horizontal) and diminish when deviated from the horizontal axis, so the lines in the horizontal axis are longer than that deviate from horizontal axis. It is clear shown in Fig.2 linear polarized laser can induce trans-azobenzene into ordered state also in the case of tightly focusing; as a result the refractive index of the azo-polymer can be changed after irradiation. It should be noted that in this case the refractive index change will be different in different directions due to the ordered arrays of the rod like azobenzene molecules.

In the next step, we try to erase the refractive index change by converting tans-azobenzene arrays into disordered state with the circular polarized 532nm laser. The calculated two-dimensional distribution of $|E|^2$ on the focal plane for a tightly focused circular polarized 532nm laser is shown in Fig.3 (a, b, c). It can be seen from the color bar that the intensity of the three components ($|E_z|^2, |E_r|^2, |E_\theta|^2$) are in the same order and should be considered in the reorientation process of the trans-azobenzene. $E_z$ is perpendicular to the azo-polymer film plane, so the long axis of the trans-azobenzene will revolve around the transverse electric vector $E_t$ ($\vec{E}_t = \vec{E}_r + \vec{E}_\theta$) and finally fall perpendicular to $\vec{E}_t$. It is clear that direction of the $\vec{E}_t$ at every point in the focal plane are not ordered, and also for circular polarized laser beam, the electric vector rotated with time, so the long axis of the trans-azobenzene will not be fixed but rotate with time. Fig.3 (d) is the temporary alignment of the trans-azobenzene under irradiation of circular polarized laser and alignment will change with time. As a result, trans-azobenzene arrays will return back to their disordered state and the refractive index change due to the irradiation of tightly focused linear polarized laser can be erased under irradiation of tightly focused circular polarized laser beam.

To test the validity of the above analyses, experiment is carried out with the set-up shown in Fig.1. Radial polarized 780nm laser is used to excite the azo-polymer film loaded surface plasmons waves propagating in all directions on the focal plane (from $0^o$ to $360^0$). If linear polarized laser beam is used to excited SPPs, the SPPs wave will only propagate in the directions near to the polarization axis, so it can not clear show the refractive index change of the azo-polymer in different directions. Fig.4 is the Fourier plane of the reflection beam (it is the image of the back focal plane of the objective) and the dark ring represents the position where the incident light converts into SPPs wave. Based on the known N.A of the objective[6], the wave-vector of the SPPs ($K_{sp}$) can be calculated. Fig.1 (a) is the image before irradiation, and the $K_{sp}$ is $1.1594K_0$, where $K_0$ is the wave-vector of the incident light in vacuum. The linear polarized 532nm laser is focused onto the azo-polymer films for three seconds then blocked up. The dark ring became larger during irradiation and stop when the 532nm laser is blocked. Fig.4 (b) is the image taken after irradiation and Fig.4 (d) are the magnified images of the areas marked with blue box , where the largest change of SPPs wave-vector ($\Delta K_{sp}$) is $0.0362K_0$, the smallest change is about $0.0201K_0$. The shifts of the points on the dark ring from the red-lines label are different from each other as shown in Fig.4(b) and (d), which is due to different refractive index change under irradiation of tightly focused linear polarized laser, and consistent with the analysis shown in Fig.2.

In the next step, one quarter wave-plate for 532nm laser is used to transfer the linear polarization to circular polarization. The circular polarized laser is focused onto the sample for another 3 second then blocked. It is evident in the experiments that the dark ring shrinks during the irradiation and stop after the laser is blocked. Fig.4 (c) is the Fourier plane of the reflection beam taken after irradiation of the tightly focused circular polarized laser. It is clearly shown that the dark ring is coincided with the red-lines label, which means that SPPs wave-vector returns back to its original value. So experimental results verify that the photo-induced refractive index change can be erased under irradiation of the tightly circular polarized laser, which is consistent with the theoretical analysis as shown in Fig.3.

The theoretical analysis and experimental results demonstrate that the wave-vector of the azo-polymer films loaded SPPs wave can be active controlled with the 532nm laser. The photo-isomerization of azobenzene is extremely rapid, occurring on picoseconds timescales, so in this experiment, the quarter wave plate can be replaced with electro-optical crystal to actively control the surface plasmons wave in high speed, which can convert the polarization state of 532nm laser in high speed. It will be done in the future work.

In summary, proof-of-concept demonstration of optical actively control of azo-polymer film loaded surface plasmons wave is presented in this paper. There are three advantages of this optical control of SPPs waves. First, there is no need to fabricate electrode on the samples as that in the electric controlling method[19, 20], so the fabrication process is easy. Second, this method allows control the dielectric loaded SPPs wave at any position of the azo-plymer films just by adjusting the samples holder stage, which is very useful in integrated optical elements, such as M-Z interferometer. Third, the spatial resolution is higher in the optical controlling method than that of electric method. In our experiments, the laser beam is expanded to full fill the whole N.A of the objective, so the radius of the focused point is about $0.61*\lambda/N.A = 228nm$. Our experimental work is useful for future development of photonic and optoelectronic integrated circuits and lab-on-a-chip sensing[8]. Future work will involve dope or synthesize the azo-molecules with photo-resist, such as PMMA for nanofabrication of integrated circuits by using Electron Beam lithography (EBL) or other lithography methods.

This work was partially supported by the National Natural Science Foundation of China under Grant No. (60778045, 60736037), the National Research Foundation of Singapore under Grant No. NRF-G-CRP 2007-01, and the Ministry of Education under Grant No. (ARC 3/06 and RGM37/06), the National Key Basic Research Program of China under Grant No. 2006CB302905. X.-C.Yuan acknowledges the support given by Nankai University (China).

**Figure Captions:**

**Fig.1:** Schematic of experimental set-up, 532nm and 780nm laser are aligned in one line and collimated before entering into the objective. Azimuth-type analyzer (AA) and spiral phase plate (SPP) for generating radial polarized light. The polarization Rotator (PR) consists of two half-wave plates.

**Fig.2:** Calculated two-dimensional distribution of $|E|^2$ for a highly focused linear polarized 532nm laser: (a, b) transversal component $|E_\theta|^2$ and $|E_r|^2$, (c) longitudinal component $|E_z|^2$, the focus is dominated by $|E_z|^2$ and $|E_r|^2$ components, (d): alignment of the trans-azobenzene according to the electric vector.

**Fig.3:** Calculated two-dimensional distribution of $|E|^2$ for a highly focused circular polarized 532nm laser: (a, b) transversal component $|E_\theta|^2$ and $|E_r|^2$, (c) longitudinal component $|E_z|^2$, the focus is dominated by three $|E_\theta|^2, |E_z|^2$ and $|E_r|^2$ components. (d): temporal alignment of the trans-azobenzene according to the electric vector.

**Fig 4:** Fourier plane of the reflected radial polarized 780nm laser beams before (a), after irradiation of the linear polarized 532nm laser (b) and subsequently irradiation of circular polarized 532nm laser (c). (d): magnified image of the part labeled with blue box shown in (b). The dark rings on the three images represent the position where SPR happens. The red lines are fixed and used to label the original position of the SPPs ring before irradiation.

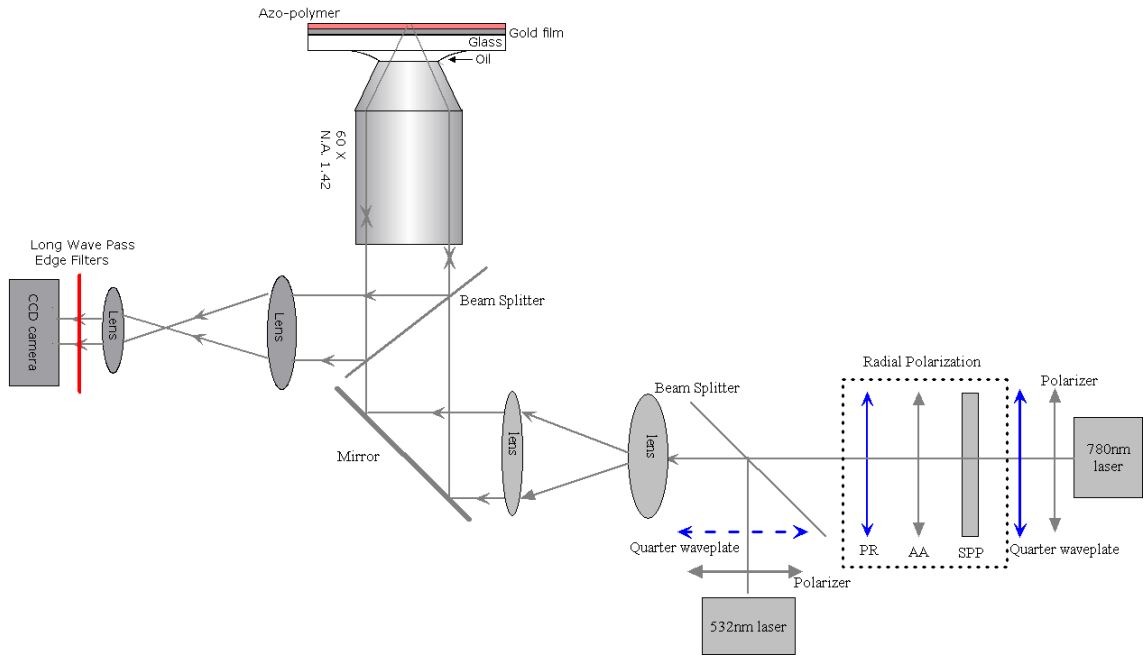

**Fig.1:** Schematic of experimental set-up, 532nm and 780nm laser are aligned in one line and collimated before entering into the objective. Azimuth-type analyzer (AA) and spiral phase plate (SPP) for generating radial polarized light .The polarization Rotator (PR) consists of two half-wave plates.

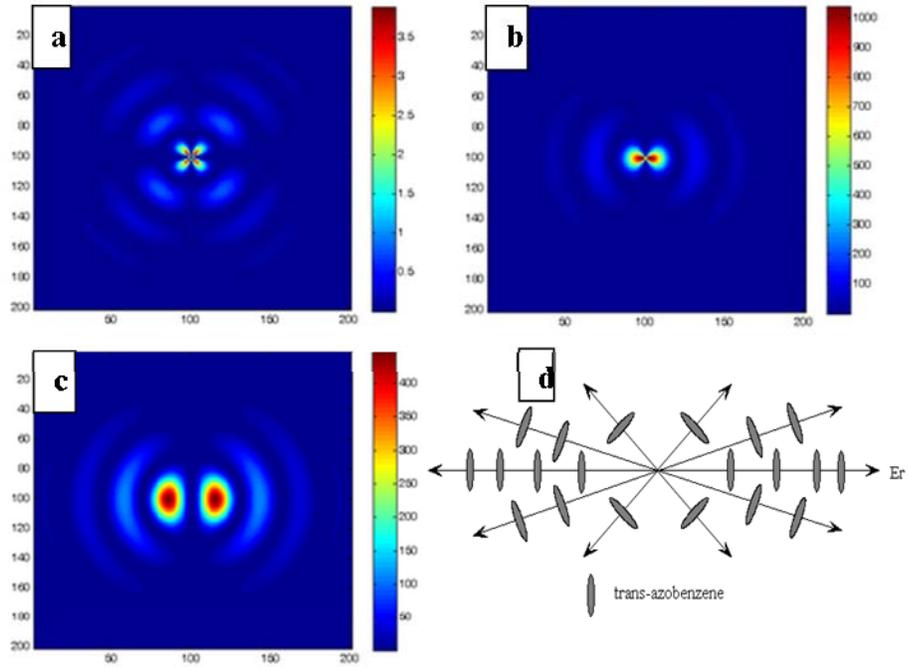

**Fig.2:** Calculated two-dimensional distribution of $|E|^2$ for a highly focused linear polarized 532nm laser: (a, b) transversal component $|E_\theta|^2$ and $|E_r|^2$, (c) longitudinal component $|E_z|^2$, the focus is dominated by $|E_z|^2$ and $|E_r|^2$ components, (d): alignment of the trans-azobenzene according to the electric vector.

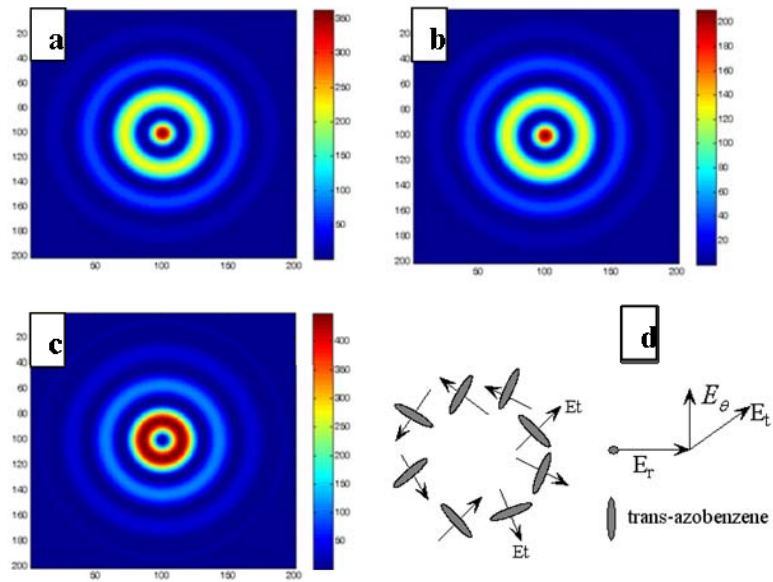

**Fig.3:** Calculated two-dimensional distribution of $|E|^2$ for a highly focused circular polarized 532nm laser: (a, b) transversal component $|E_\theta|^2$ and $|E_r|^2$, (c) longitudinal component $|E_z|^2$, the focus is dominated by three $|E_\theta|^2, |E_z|^2$ and $|E_r|^2$ components. (d): temporal alignment of the trans-azobenzene according to the electric vector.

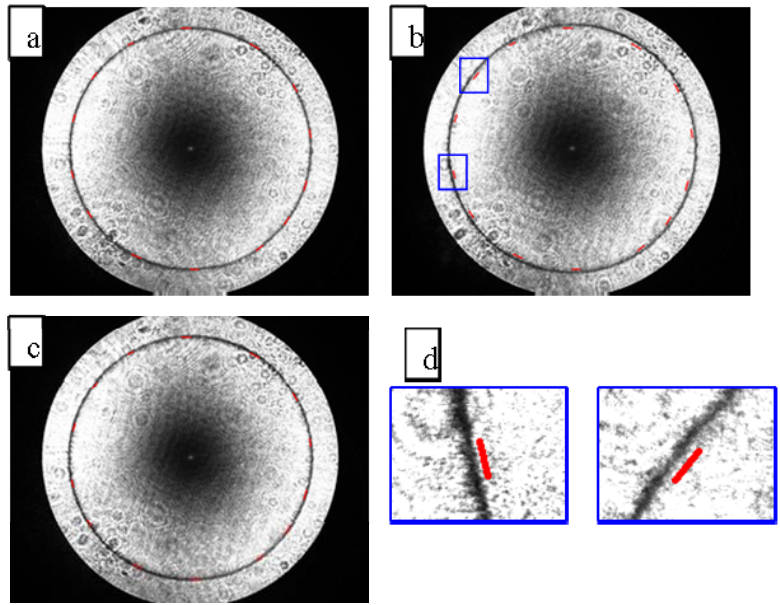

**Fig 4:** Fourier plane of the reflected radial polarized 780nm laser beams before (a), after irradiation of the linear polarized 532nm laser (b) and subsequently irradiation of circular polarized 532nm laser (c). (d): magnified image of the part labeled with blue box shown in (b). The dark rings on the three images represent the position where SPR happens. The red lines are fixed and used to label the original position of the SPPs ring before irradiation.